\input{aipcheck}

\documentclass[prd,amsmath,amssymb,letterpaper,english]{aipproc}
\usepackage{babel}
\usepackage{amsmath}
\usepackage{wrapfig}
\usepackage{graphicx}
\usepackage{latexsym}
\usepackage{epsfig}
\layoutstyle{6x9}
\usepackage{epsfig}

\def \snu   {\ensuremath{ \tilde{\nu}               }}

\def \sle   {\ensuremath{ \tilde{\ell}              }}

\def \smu   {\ensuremath{ \tilde{\mu}               }}
\def \sta   {\ensuremath{ \tilde{\tau}              }}

\def \squark {\ensuremath{ \tilde{q}                 }}
\def \gluino {\ensuremath{ \tilde{g}                 }}
\def \stopstop  {\ensuremath{ \tilde{t}                 }}
\def \stopquark {\ensuremath{ \tilde{t}_\mathrm{1}      }}
\def \stop2 {\ensuremath{ \tilde{t}_\mathrm{2}      }}

\def \lep   {\ensuremath{ \ell                      }}

\def \mstop  {\ensuremath{ m_{\tilde{\mathrm{t}}}  }}
\def \msquark  {\ensuremath{ m_{\tilde{\mathrm{q}}}  }}
\def \mstop1 {\ensuremath{ m_{\tilde{\mathrm{t1}}}  }}

\newcommand{\Rslash}{\mbox{$R \kern-0.6em\slash$}}
\newcommand{\Rpslash}{\mbox{$\Rslash_{p}$}}
\newcommand{\Eslash}{\mbox{$E \kern-0.6em\slash$ }}
\newcommand{\etmiss}{\mbox{$\Eslash_{T}\!$     }}

\begin{document}

\title{SUSY Multilepton Signatures at Tevatron\footnote{Talk presented
at the Aspen Winter Conference ``New Physics at the Electroweak Scale
and New Signals at Hadron Colliders, January 8-14 (2007).}}


\author{Maxim Titov$^*$}{
  address={Albert-Ludwigs University of Freiburg, Physics Institute,\\
79104, Freiburg, Germany\\
(on behalf of the CDF and D\O~Collaborations)\\
$^*$E-mail: maxim.titov@physik.uni-freiburg.de\\
}
}



\begin{abstract}


 One of the most striking signature of supersymmetric models with
electroweak symmetry breaking is the presence of 
multilepton event topologies in the 
decay products.
 In this paper searches are presented for physics beyond the Standard Model (SM) 
in final states containing charged leptons
from proton-antiproton collision data at a center-of-mass energy of 1.96~TeV,
collected with Run~II CDF and D\O~Detectors in 2002-2006,  and
corresponding to integrated luminosities of up to 1.1~fb$^{-1}$.
 In any of the searches
no excess of candidates was observed with respect
to the SM predictions and limits on masses and production
cross-sections are set at the 95~$\%$ CL.



\end{abstract}

\maketitle


\section{Introduction}

 Supersymmetric extensions of the Standard Model (SM) provide a consistent framework for
gauge unification and stabilisation of electroweak scale~\cite{martin}.
 Within the minimal supersymmetric extension of the SM model (MSSM) the stability of the 
Lightest Supersymmetric Particle (LSP) is ensured by the multiplicatively
conserved quantum number $R = (-1)^{3B+L+2S} = 1 (-1)$
for ordinary particles (sparticles), where
$B,L,S$ are the baryon, lepton and spin quantum numbers, respectively.
 This assumption has phenomenological consequences: 
SUSY particles are only pair produced
and the LSP interacts weakly with ordinary matter.
 The description of electroweak symmetry breaking requires more than 100 parameters,
which can be significantly reduced in certain models.
 A minimal supergravity model (mSUGRA) with $R$-parity ($R_p$) conservation,
which is the framework for many experimental searches at Tevatron,
has only 5 free parameters:
$m_0$, $m_{1/2}$ are the common scalar (Higgs, sleptons and squarks) and
gaugino (bino, wino and gluino) masses, respectively,
$A_0$, the trilinear scalar coupling; $\tan \beta$, the ratio of vev's 
of two Higgs doublets; sign($\mu$), the Higgs sector mass parameter.

However, neither gauge invariance nor SUSY require $R_p$ conservation.
 Allowing $R$-parity violation requires 45 extra parameters.
 The most general superpotential $W = W_{MSSM} + W_{RPV}$ contains 
explicit $R$-parity violating trilinear ($\Rpslash$),
allowing the LSP to decay into SM particles, and 
bilinear ($B \Rpslash$) terms~\cite{barbier}:
\begin{equation}
W_{RPV}= \frac{1}{2} \lambda_{ijk} L_i L_j \overline{E_k} +
 \lambda_{ijk}^{'} L_i Q_j \overline{D_k} + 
 \frac{1}{2} \lambda_{ijk}^{''} \overline{U_i}~\overline{D_j}~\overline{D_k} +
 \mu_i L_i H_u
\end{equation}
 where $L$ and $Q$ are left-handed lepton and quark $SU(2)$ doublet superfields,
$E, U$ and $D$ are right-handed lepton and quark singlet superfields 
and $i,j,k$ indicate flavour indices.
 The last term describes the bilinear interactions between left-handed lepton and Higgs superfield.
 It is also assumed that $\Rpslash$ couplings are small 
compared to gauge strength and have a negligible effect on the
Renormalization Group Equations (RGE) and on the running of the 
soft-supersymmetry breaking parameters and lepton Yukawa coupling.


 An overview of several key multilepton signatures at Tevatron, followed
by the experimental results of SUSY~searches both in $R_p$-conserving and $\Rpslash$
scenarios are presented in this paper. Finally, ``signature-based'' searches are discussed.

\section{Multilepton Signals from Supersymmetry}

 Leptons represent the most powerful discriminating signatures at hadron colliders.
 Despite the fact that leptonic branching ratios are not always favourable,
multilepton signals of new physics can become competitive to the SM 
backgrounds for a wide range of the SUSY parameter space of masses and couplings.
In the following, we would like to discuss few distinctive multilepton signatures,
which might yield evidence for supersymmetric particle production at Tevatron,
as luminosity increases.

\begin{table}[htb]
\begin{tabular}{ccc}
\hline
 {\bf Signature} & {\bf Production} & {\bf Decay} \\
\hline
\multicolumn{3}{c}{\bfseries $R$-Parity ($R_p$) Conserving Signatures $\rightarrow$ LSP~(${\tilde \chi_1^0}$) is stable} \\
\hline
$2 \lep + \etmiss $    & ${\tilde \chi_1^{\pm}} {\tilde \chi_1^{\mp}}$ 
                       & ${\tilde \chi_1^{\pm}} \rightarrow {\tilde \chi_1^0} \lep^{\pm} \nu$ \\
                       & ${\tilde \chi_2^0} {\tilde \chi_2^0}$ & ${\tilde \chi_2^0} \rightarrow {\tilde \chi_1^0} \lep \lep$, 
                         ${\tilde \chi_2^0} \rightarrow {\tilde \chi_1^0} \nu \nu $ \\
                       & $\sle \sle$ & $\sle \rightarrow {\tilde \chi_1^0} \lep^{\pm}$, 
                         $\sle \rightarrow {\tilde \chi_1^{\pm}} \nu$, 
                         ${\tilde \chi_1^{\pm}} \rightarrow {\tilde \chi_1^0} \lep^{\pm} \nu $\\
$\ge 3 \lep + \etmiss $& {\bf {${\tilde \chi_1^{\pm}} {\tilde \chi_2^0}$}} & 
                         {\bf {${\tilde \chi_1^{\pm}} \rightarrow {\tilde \chi_1^0} \lep^{\pm} \nu$}},
                         {\bf {${\tilde \chi_2^0} \rightarrow {\tilde \chi_1^0} \lep \lep $}} \\
                       & ${\tilde \chi_2^0} {\tilde \chi_2^0}$ & ${\tilde \chi_2^0} \rightarrow {\tilde \chi_1^0} \lep \lep$ \\
                       & $\sle \sle$ & $\sle \rightarrow {\tilde \chi_2^0} \lep^{\pm}$,
                         ${\tilde \chi_2^0} \rightarrow {\tilde \chi_1^0} \lep \lep$ \\ 
                       & $\sle \snu$ & $\sle \rightarrow {\tilde \chi_2^0} \lep^{\pm}$,
            ${\tilde \chi_2^0} \rightarrow {\tilde \chi_1^0} \lep \lep$, 
            $\snu \rightarrow {\tilde \chi_1^{\pm}} \lep^{\mp}$ \\ 
$2 \lep + \ge$~1~jet~$+ \etmiss $  &  ${\tilde \chi_1^{\pm}} {\tilde \chi_2^0}$ 
                                 &  ${\tilde \chi_1^{\pm}} \rightarrow {\tilde \chi_1^0} q q^{'}$,
                                    ${\tilde \chi_2^0} \rightarrow {\tilde \chi_1^0} \lep \lep $ \\
                                 &  $\gluino \gluino, \squark \squark $
                                 &  $\gluino \rightarrow {\tilde \chi_1^{\pm}} q q^{'}$,
                                    $\squark \rightarrow {\tilde \chi_1^{\pm}} q^{'}$\\
                                 &  $\stopquark \stopquark$
                                 &  $\stopquark \rightarrow b {\tilde \chi_1^{\pm}}$, 
                                    ${\tilde \chi_1^{\pm}} \rightarrow {\tilde \chi_1^0} \lep^{\pm} \nu$ \\
$\ge 3 \lep + \ge$~1~jet~ $\etmiss $& {\bf {${\tilde \chi_1^{\pm}} \chi_3^0$}} & 
                                    ${\tilde \chi_1^{\pm}} \rightarrow {\tilde \chi_1^0} \lep^{\pm} \nu$,
                                    $\chi_3^0 \rightarrow {\tilde \chi_2^0} q q $ \\
                                 &  $\gluino \gluino, \squark \squark, \gluino \squark$
                                 &  $\gluino \rightarrow {\tilde \chi_1^{\pm}} q q^{'}$,
                                    $\gluino \rightarrow {\tilde \chi_2^0} q q $,
                                    $\squark \rightarrow {\tilde \chi_2^0} q$\\
\hline
\multicolumn{3}{c}{\bfseries $R$-parity Violating ($\Rpslash$) Signatures $\rightarrow$ LSP~(${\tilde \chi_1^0}$) decays to SM particles}\\
\hline
$\ge 2 \lep + \ge 0 jet + \etmiss $ & $\sle, \snu$ &
                                      $\sle \rightarrow \lep {\tilde \chi_1^0}, {\tilde \chi_1^0} \rightarrow q q^{'} \lep$,
                                      $\snu \rightarrow \lep \lep $\\
$\ge 3 \lep + \ge 0 jet + \etmiss $ ~~~~~~~~~& ${\tilde \chi_1^0} {\tilde \chi_1^0}, {\tilde \chi_1^{\pm}} {\tilde \chi_2^0}, 
                                       {\tilde \chi_1^{\pm}} {\tilde \chi_1^{\mp}}, {\tilde \chi_2^0} {\tilde \chi_2^0}$ 
                                    ~~~~~~~~~& ${\tilde \chi_1^0} \rightarrow \lep \lep \nu$,
                                      ${\tilde \chi_1^{\pm}} \rightarrow {\tilde \chi_1^0} \lep^{\pm} \nu$,
                                      ${\tilde \chi_2^0} \rightarrow {\tilde \chi_1^0} \lep \lep $\\

\hline
\end{tabular}
\caption{ Example of SUSY Multilepton Signatures at Tevatron
in $R$-parity ($R_p$) Conserving and R-parity Violating ($\Rpslash$) Scenarios. 
Different decay modes between the $\sle_L$, $\sle_R$ and $\squark_L$, $\squark_R$ 
are not shown.}
\label{summary:multileptondecays}
\end{table}

{\bf {Charginos and Neutralinos.}}
 At Tevatron, charginos and neutralinos can be pair produced 
via their electroweak couplings to the gauge bosons ($\gamma,W,Z$)
in the $s$-channel and to squarks in the ($t,u$)-channel
of $q \overline {q}$ annihilation (see Fig.~\ref{feynman:rpcsusy}).
 The production cross section is not only a function
of chargino and neutralino masses but depends strongly
on the gaugino-higgsino model-dependent mixing and the squark masses.
In most regions of mSUGRA parameter space, 
masses of $\chi_{1,2}^0$ and ${\tilde \chi_1^{\pm}}$ and gluino behave gaugino-like, 
i.e depend on $m_{1/2}$. 
Analytical expressions for the squark ($\squark$) and slepton ($\sle$) mass
parameters can be obtained when the corresponding Yukawa couplings are negligible.
For small values of $\tan \beta$, masses of $\squark$ and $\sle$
depend both on $m_0$ and $m_{1/2}$.

\setlength{\unitlength}{1mm}
\begin{figure}[bth]
 \begin{picture}(22,22)
\put(-65.0,-6.0){\includegraphics{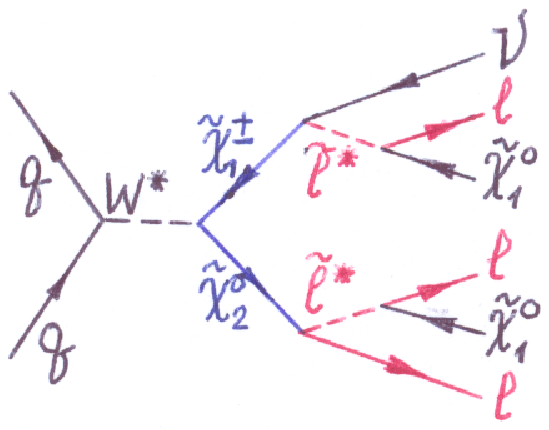}}
\put(-28.0,-6.0){\includegraphics{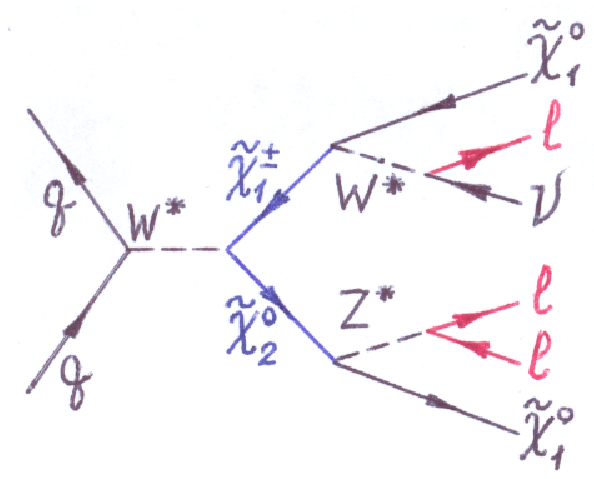}}
\put(9.0,-6.0){\includegraphics{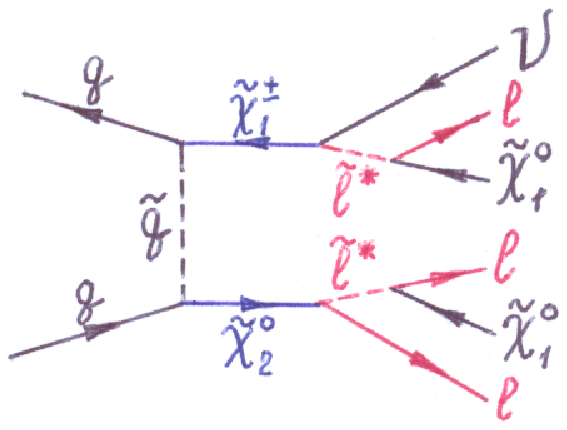}}
\put(46.0,-7.0){\includegraphics{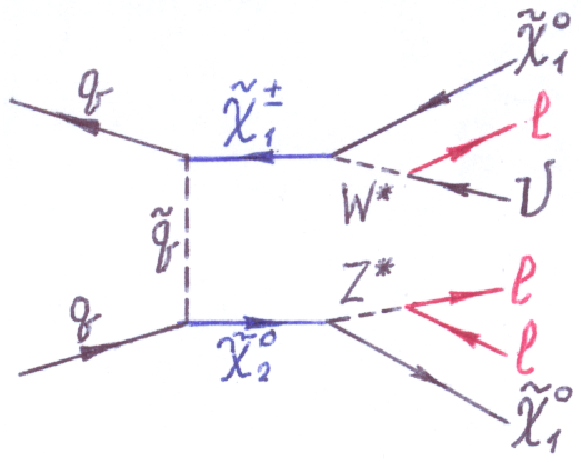}}
\put(-61.0,22.0){\footnotesize \bf a)}
\put(-24.0,22.0){\footnotesize \bf b)}
\put(13.0,22.0){\footnotesize \bf c)}
\put(50.0,22.0){\footnotesize \bf d)}
 \end{picture}
\caption{\footnotesize
 Feynman diagrams for associated ${\tilde \chi_2^0} {\tilde \chi_1^{\pm}}$ production
at leading order in electroweak $s$-channel (a,b) 
and $t$-channel (c,d) reactions of a quark and anti-quark
and subsequent decays via sleptons (a,c) or via vector bosons (b,d).
The $\chi_{1}^0$ is assumed to be the LSP in mSUGRA.
\label{feynman:rpcsusy}}
\end{figure}

 In $R_p$-conserving scenario, one of the most promising channels
for SUSY searches is the trilepton signature arising from the
chargino-neutralino production ($p \overline{p} \rightarrow {\tilde \chi_1^{\pm}} {\tilde \chi_2^0}$) 
with subsequent leptonic decays
(${\tilde \chi_1^{\pm}} \rightarrow {\tilde \chi_1^0} \lep \nu$,
${\tilde \chi_2^0} \rightarrow {\tilde \chi_1^0} \lep \lep$).
 The overall ${\tilde \chi_1^{\pm}} {\tilde \chi_2^0} \rightarrow 3\lep$ detection efficiency 
depends on the relative contribution from the $\sle$ and $W/Z$-exchange graphs, 
which varies as a function of $m_{\sle}$.  
 At large values of $m_0$ ($m_0 \sim 1$~TeV) sleptons are heavy,
thus slepton mediated diagrams of ${\tilde \chi_2^0}/{\tilde \chi_1^{\pm}}$ are suppressed, and gauginos
decay to trilepton final states dominantly 
via real (at large $m_{1/2}$) or virtual (small $m_{1/2}$) $W/Z$-boson exchange
(see Fig.~\ref{feynman:rpcsusy} (b,d)).
 As the $m_0$ decreases, the slepton masses also decrease and the
${\tilde \chi_1^{\pm}}/{\tilde \chi_2^0}$ decays via virtual sleptons 
(3-body decay ${\tilde \chi_2^0} \rightarrow \sle^{*} \lep \rightarrow {\tilde \chi_1^0} \lep \lep$)
become maximally enhanced at $m_0 \sim 100$~GeV ($m_{1/2} \sim 200$~GeV).
 For even lower values of $m_0$ ($m_0 \le m_{1/2}/2$) sleptons are
lighter than ${\tilde \chi_1^{\pm}}/{\tilde \chi_2^0}$ and gaugino dominantly decays to the 
on-shell sleptons (2 body decay ${\tilde \chi_2^0} \rightarrow \lep \sle$),
as shown in Fig.~\ref{feynman:rpcsusy} (a,c).
 If charginos ($\tilde \chi_{1,2}^{\pm}$) and neutralinos 
($\tilde \chi_{2,3,4}^{0}$) are not too heavy (low $m_{1/2}$) 
they can contribute to the multilepton
signatures (at low $m_0$) via $\tilde \chi_{2}^{0} \tilde \chi_{2}^{0}, \tilde \chi_{2}^{0} \tilde \chi_{3}^{0},
\tilde \chi_{1}^{\pm} \tilde \chi_{3}^{0}, \tilde \chi_{2}^{\pm} \tilde \chi_{3}^{0}$ production 
(see Table~\ref{summary:multileptondecays}).

\setlength{\unitlength}{1mm}
\begin{figure}[bth]
\begin{picture}(24,24)
\put(-60.0,-6.0){\includegraphics{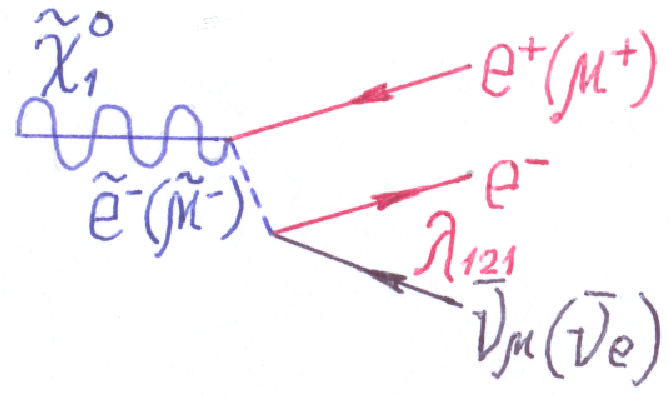}}
\put(-15.0,-6.0){\includegraphics{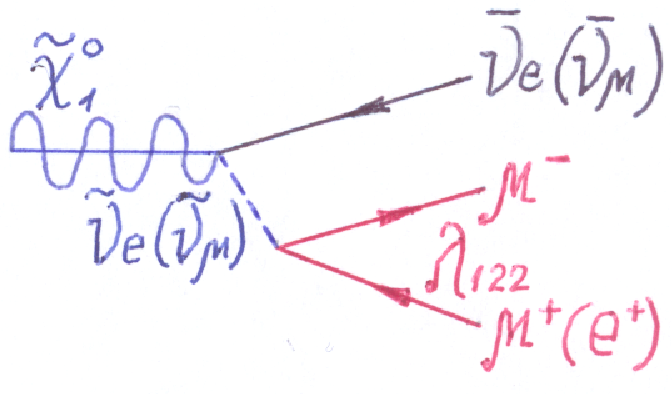}}
\put(35.0,-6.0){\includegraphics{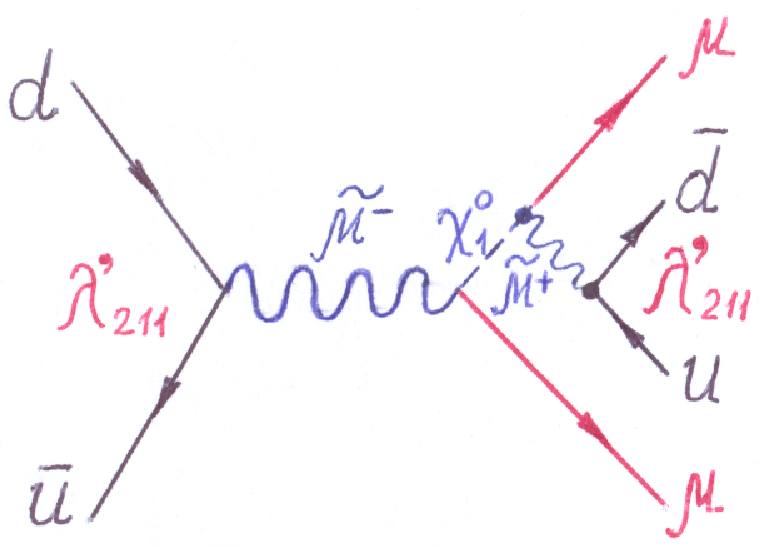}}
\put(-40.0,22.0){\footnotesize \bf a)}
\put(5.0,22.0){\footnotesize \bf b)}
\put(50.0,22.0){\footnotesize \bf c)}
\end{picture}
\caption{\footnotesize
Representative diagrams for the $R$-parity violating ($\Rpslash$) decays of ${\tilde \chi_1^0}$
(assumed as LSP) via $LL\overline{E}$ couplings: (a) $\lambda_{121}$ and (b) $\lambda_{122}$.
Example of resonant smuon production and decay via $\lambda^{'}_{211} L Q \overline{D}$ coupling (c).
\label{feynman:rpvsusy}}
\end{figure}

 The popular $R_p$-conserving scenario assumes that the LSP is stable
and neutral (for cosmological reasons) and therefore escapes detection in the experiment.
 This picture changes dramatically with $\Rpslash$.
 The LSP is no longer stable and carries away transverse energy, and the 
$\etmiss$ signal, which is the mainstay of SUSY searches in $R_p$-conserving
models, is greatly degraded.
 A fundamental $\Rpslash$ interaction could mediate the decay of any 
chargino or neutralino\footnote{
For sparticles other than the LSP, the $\Rpslash$ decays will in general
compete with $R_p$-conserving ``cascade decays'' initiated by gauge interactions.},
as illustrated in Fig.~\ref{feynman:rpvsusy} (a,b) for ${\tilde \chi_1^0}$,
thus converting the $R_p$ signal ${\tilde \chi_1^{\pm}} {\tilde \chi_2^0} \rightarrow 3 \lep$
into final state with up to seven charged leptons.
 Even when $R_p$ conserving leptonic decays of ${\tilde \chi_1^{\pm}} {\tilde \chi_2^0}$ are suppressed
at least 4 leptons can still be present in $\Rpslash$-decay
via $\lambda$-coupling.
 Furthermore, $\Rpslash$ multilepton signals can also appear from 
${\tilde \chi_1^{\pm}} ({\tilde \chi_2^0} \rightarrow {\tilde \chi_1^0} \nu \nu), {\tilde \chi_1^{\pm}} {\tilde \chi_1^0},
{\tilde \chi_1^{\pm}}  \chi_1^{\mp}$ channels, that give no $R_p$ trileptons.
  The actual multiplicity of observed leptons depends on the exact $\Rpslash$
scenario\footnote{ The results are very model dependent, owing to the large
parameter space in the $\Rpslash$ sector.
 While LSP decays via $L$-violating $\Rpslash$ ($\lambda$, $\lambda^{'}$)-terms
give rise to multilepton and lepton+jets final states,
the $\chi_1^0$ decays via $\Rpslash$ $\lambda^{''}$-coupling lead to 
the QCD multijet final states,
difficult to access at hadron colliders.
}, 
the experimental thresholds and angular acceptances,
in general, leading to high visibility of the multilepton signals.
 The presence of bilinear $B\Rpslash$ coupling can generate neutrino masses,
in agreement with the atmospheric neutrino data, and to enhance SUSY multilepton signals,
especially at moderate and large $m_0$~\cite{magro}.



{\bf{Stop Quarks.}} 
 Squarks of the $3^{rd}$ generation deserve a special attention, since
 due to the impact of the large top Yukawa coupling on the RGE
 the stop mass drops to the lowest value in the squark spectrum.
 Moreover, the top quark eigenstates in the weak bases $\stopstop_L$ and $\stopstop_R$
will strongly mix due to the large $m_t$, leading to a small mass value of $\stopquark $
possibly much lighter than other squarks.
 At hadron colliders, the cross sections for the production of a light scalar top pairs
$gg/q\overline{q} \rightarrow \stopquark \overline{\stopquark }$
depend essentially only on the $m_{\stopquark }$ and very little 
on the other SUSY parameters (i.e. $m_{\gluino}$, $m_{\squark}$ and
the mixing angle in the top squark sector), which affect only the higher-order corrections.
The Tevatron lower limits on $m_{\squark}$ are derived assuming ten 
degenerate squark flavours and not applicable to~$\stopquark $.

\setlength{\unitlength}{1mm}
\begin{figure}[bth]
\begin{picture}(35,35)
\put(-50.0,-6.0){\includegraphics{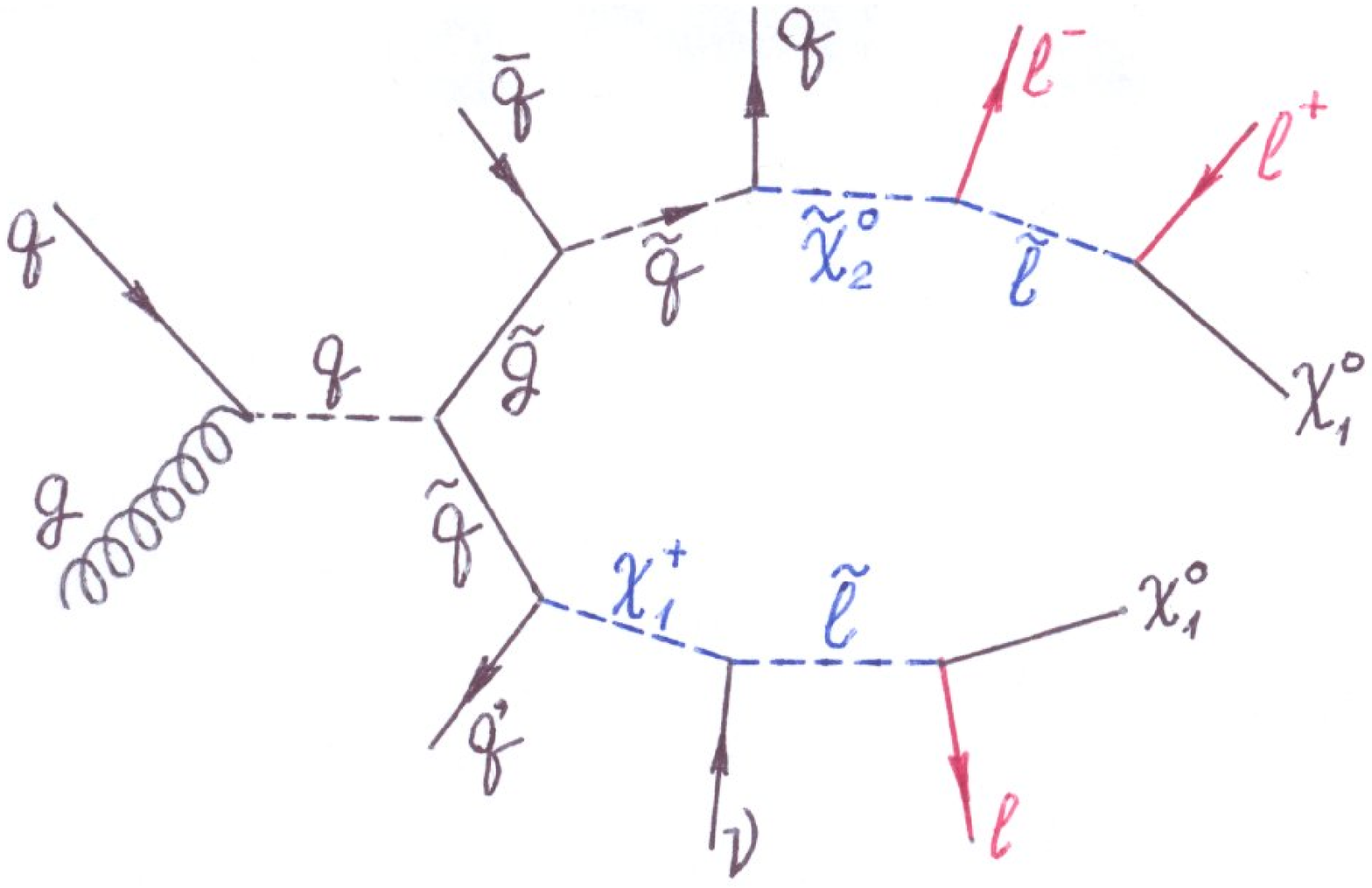}}
\put(20.0,-5.0){\includegraphics{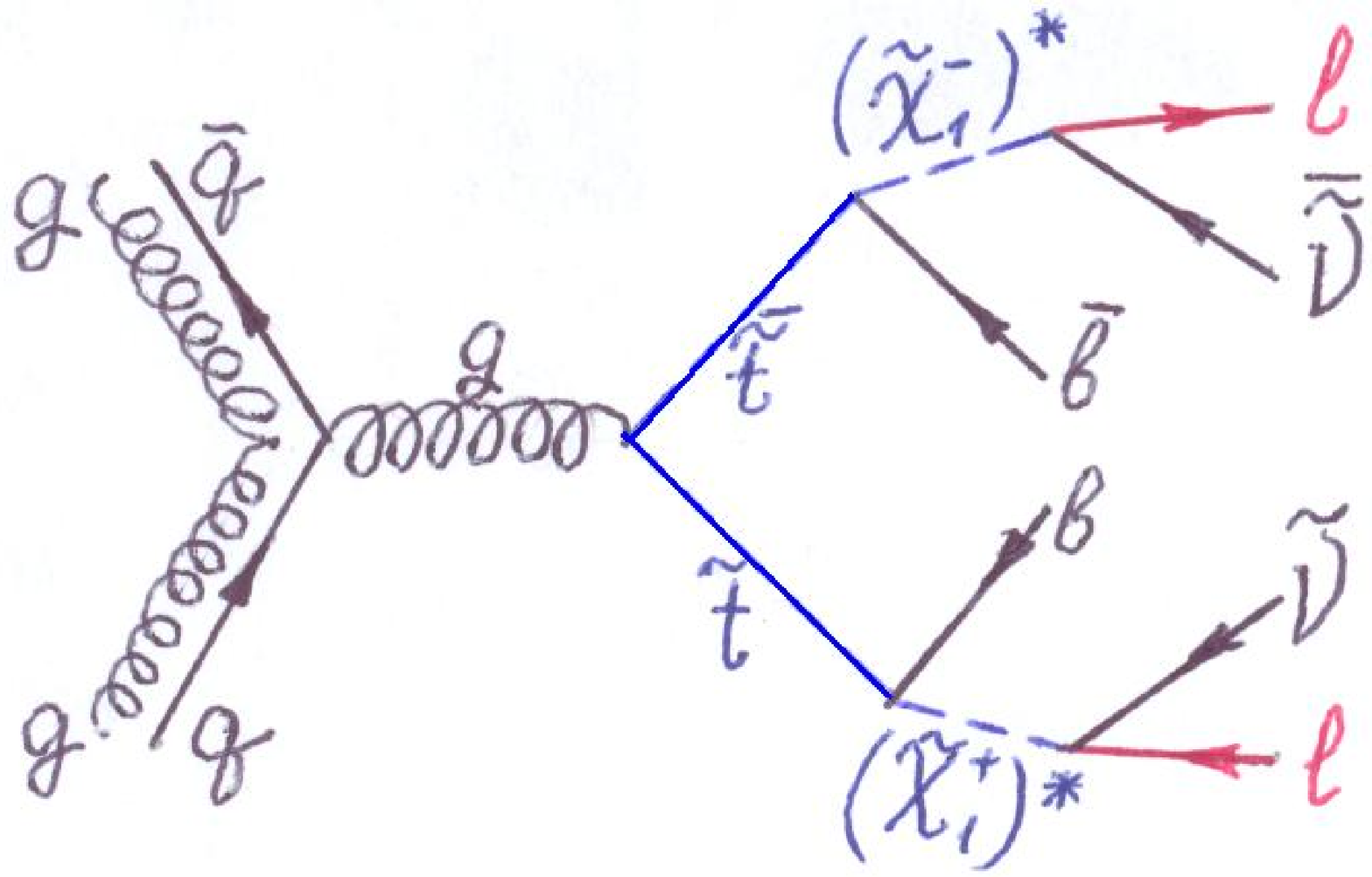}}
\put(-40.0,30.0){\footnotesize \bf a)}
\put(20.0,30.0){\footnotesize \bf b)}
\end{picture}
\caption{\footnotesize
(a) Feynman diagram for squark/gluino production with subsequent decays
leading to 3 leptons, jets and $\etmiss$ final state. 
(b) Example of the $\stopquark $ decay to lepton and sneutrino, via 
virtual gaugino. 
When the 3-body
decay into $b \lep \snu$ is kinematically allowed, the subsequent invisible decay of
$\snu \rightarrow \nu {\tilde \chi_1^0}$ has no influence on kinematics.
\label{feynman:squarkgluino}}
\end{figure}

 The signals from top squark $\stopquark $ at colliders depend on its decay pattern.
 The most explored signature 
for a light stop $\stopquark $ production at Tevatron
is a loop-induced and flavor-changing decay: $\stopquark \rightarrow c {\tilde \chi_1^0}$, 
when the main tree-level diagrams $\stopquark \rightarrow b {\tilde \chi_1^{\pm}}$  
($\stopquark \rightarrow t {\tilde \chi_1^0}$) for medium (heavy) stop quarks
are kinematically forbidden~\cite{cdfstop}.
 However, recent studies also favour 3-body decays 
$\stopquark \rightarrow t^{*} (\chi_1^{0}) \rightarrow b W {\tilde \chi_1^0}$ and
$\stopquark \rightarrow b ({\tilde \chi_1^{\pm}})^{*} \rightarrow b \sle \nu, b \lep \snu$, 
and, if kinematically not allowed, even 4-body decay
$\stopquark \rightarrow b \lep \nu {\tilde \chi_1^0}$~\cite{scalartop}.
 The later mode is mediated by virtual chargino and is the same order of
perturbation theory as the loop-induced $\stopquark \rightarrow c {\tilde \chi_1^0}$ decay,
i.e $O (\alpha^3)$.
 The 3-body $\stopquark \rightarrow b \lep \snu$ decay
might be favoured over $\stopquark \rightarrow b W {\tilde \chi_1^0}$
and $\stopquark \rightarrow b \sle \nu$\footnote{
At small $\tan \beta$, decays into sneutrino are more important 
than into charged sleptons as a result of different
spin structure of corresponding matrix elements~\cite{porod}.}
due to the relatively weak constraint 
on the sneutrino mass at LEP $m_{\snu} >$ 45~GeV~\cite{lepsusy}.
 The experimental signature for the 3-body $\stopquark \overline{\stopquark}$ decay,
via virtual gauginos, is $2\lep$+$2b$-jets+$\etmiss$ (see Fig.~\ref{feynman:squarkgluino} (b)),
assuming $R$-parity conservation.
 This mode can even dominate at Tevatron for $m_{\stopquark} < 200$~GeV
if charginos and sleptons masses are not too much larger than
their present experimental bounds~\cite{porod}.

{\bf {Gluinos and Squarks.}}
 At Tevatron, heavy gluinos ($\gluino$) and squarks ($\squark$),
produced via their $SU(3)_c$ coupling to quarks and gluons,
tend to decay via sequence
of cascades through charginos (${\tilde \chi_1^{\pm}}$) and neutralinos (${\tilde \chi_2^0}$)
modes which, unless suppressed by phase space, frequently dominate 
the $R_p$-conserving decays of $\squark_{L}$ and $\gluino$. 
 The subsequent leptonic decays of ${\tilde \chi_1^{\pm}}$ and ${\tilde \chi_2^0}$
yield events with hard jets accompanied by 1-3 isolated leptons 
and $\etmiss$ (see Fig.~\ref{feynman:squarkgluino} (a)).
 The kinematics of signal events is usually harder 
than that of SM backgrounds for large squark-gluino 
masses\footnote{The $p_T$ of primary jets and $\etmiss$
are expected to scale with $m_{\gluino}$ and $m_{\squark}$.}. 
In contrast, the difference in the lepton momenta might be not very pronounced
as signal leptons are produced far down in cascade decay chain, 
thus loosing ``memory'' about the hardness of the original process. 
 While there are substantial backgrounds from $W$+jet and QCD multijet
to $\lep$ topology,
the physics backgrounds in the 
$\lep^+ \lep^-$, $\lep^{\pm} \lep^{\pm}$ \footnote{Gluino pair production is a copious source
of like-sign dilepton pairs. Since gluino is a Majorana particle it 
decays with the same probability into chargino of either sign.}
and $3 \lep$ channels can be better controlled, so that these
searches are essentially rate limited.
 The experimental signature ($\ge 2\lep$)+jets+$\etmiss$ provide a
complimentary ways to search for gluinos and squarks at Tevatron
to the classic $\etmiss$+multijet signatures.

{\bf{Sleptons}}. Charged sleptons and sneutrinos can be pair produced at hadron
colliders via the Drell-Yan mechanism $p \overline{p} \rightarrow Z^*/\gamma^* \rightarrow
\sle\sle$ and can be detected through the slepton decays $\sle \rightarrow \lep {\tilde \chi_1^0}$
in a clean $\lep^+ \lep^-$ + $\etmiss$ final states.
 This channel have received rather limited attention at Tevatron
due to the smallness of cross section: $\sigma (\sle\sle) \sim 10-50$~fb 
for $m_{\sle} \sim 100$ GeV.
 Untangling a few slepton events from the major
$WW, Z/\gamma^*$
would be very difficult, since the
SM contributions is at least few times larger than the expected SUSY signal.
 This makes unlikely for Tevatron to detect sleptons in $R_p$ conserving 
scenarios beyond the range of LEP~\cite{lepsusy}.
 In $\Rpslash$ models, slepton and sneutrino resonant production 
occurs through the new $\lambda_{ijk}^{'}$ interactions. 
 Tevatron data allows a search for a single slepton production over a considerable
range of masses and for $\lambda_{ijk}^{'}$ values down to $10^{-2}$ 
(see Fig.~\ref{feynman:rpvsusy} (c)).

{\bf{Signature-based Searches}}.
 Motivated by the detection of the CDF Run I $ee\gamma\gamma\etmiss$
event (on a background of $<10^{-6}$), a $\mu\mu\gamma\gamma jj$
event (on a background of $< 10^{-5}$) and 2.7~sigma excess above the SM 
predictions in the $\lep\gamma\etmiss$ final state~\cite{cdf:lgamma}, 
``signature-based'' inclusive searches 
became an important experimental method at Tevatron. 
 In particular, model-independent anomalous production of 
$\lep\gamma$+X ($\lep\gamma\etmiss, \gamma\gamma\etmiss, \lep\gamma\gamma, \lep\lep\gamma$)
events, expected in gause-mediated models of supersymmetry or
in the production of excited leptons,
 can be sensitive to any new physics beyond the SM.






\section{SUSY Searches with R-parity conservation}


 Both CDF and D\O~have exploited the classical SUSY trilepton signature 
coming from $p \overline{p} \rightarrow {\tilde \chi_1^{\pm}} {\tilde \chi_2^0}$ pair production
followed by ${\tilde \chi_1^{\pm}} \rightarrow {\tilde \chi_1^0} \lep^{\pm} \nu$
and ${\tilde \chi_2^0} \rightarrow {\tilde \chi_1^0} \lep \lep$ decays, with hadronic activity
coming only from QCD radiation.  
 Common challenges for analysis are the estimation of the third
lepton misidentification probability and conversion modelling. 

 D\O~has searched for the $3 \lep + \etmiss$ final state in 6
different analyses, including those with hadronic $\tau$ decays
($\tau_{had}$),
listed in Table~\ref{d0:trileptonsummary}~\cite{prltrilepton,d0note5127}.
 While very few SM processes contribute to the trilepton signature, 
the leptons from ${\tilde \chi_1^{\pm}} {\tilde \chi_2^0}$ can be very soft,
requiring efficient lepton identification and 
multilepton triggering at low $p_T$\footnote{
D\O~uses combination of single lepton, lepton+track and dilepton triggers to select data events.}. 
 The analysis strategy is to require two isolated leptons ($\lep = e, \mu, \tau_{had}$)
with $p_T^{\lep 1~(\lep 2)} > 12~(8)$~GeV
satisfying analysis dependent topological cuts and the presence of large missing 
$\etmiss$ in the event.
 In all the channels the dilepton invariant mass must be in excess of 15-25 GeV 
to suppress QCD, $J/\psi$, $\Upsilon$ and $W\gamma^*$ contributions
and be away from the $Z$-mass region.
$t\overline{t}$ background is discriminated by vetoing events with 
large jet activity, while $b\overline{b}/c\overline{c}$ leptons
carry low $p_T$ and appear non-isolated in calorimeter.
 To minimise lepton identification inefficiencies at low $p_T$,
the third lepton is reconstructed
as an isolated, high-quality track with $p_T^{\lep 3} > 3-5$~GeV 
originating from the same vertex as the two identified leptons
\footnote{Relaxing the third lepton requirement to an isolated
low-$p_T$ track also increases the sensitivity of the search to $\tau$ lepton
decays, whose contribution becomes dominant at large $\tan \beta$ scenarios.}.
 The track (calorimeter) isolation 
for the third track is designed to be efficient
for all lepton flavours
($\lep = e, \mu, \tau_{had}$).
 Alternatively, the three-lepton mode is relaxed to a same-sign 
$\mu^{\pm} \mu^{\pm}$ selection, where SM background is low enough.
The dominant irreducible background 
consists of $WZ/ZZ$ associated production with subsequent leptonic decays of gauge bosons.
 The major SM backgrounds remaining after selection cuts are due to
W+jet/$\gamma$, $Z/\gamma^{*}$ and $WW$ pair production, where
the light $(u,d,s)$-quark jets and/or photon
conversions are misidentified as leptons.

\begin{table}[htb]
\begin{tabular}{lllc|lllc}
\hline
    \tablehead{1}{c}{b}{D\O~Analysis}
  & \tablehead{1}{c}{b}{$ L_{int}$ \\(fb$^{-1})$}
  & \tablehead{1}{c}{b}{Data}
  & \tablehead{1}{c}{b}{SM bkg}
\vline
  & \tablehead{1}{c}{b}{CDF~Analysis}
  & \tablehead{1}{c}{b}{$ L_{int}$ \\(fb$^{-1})$}
  & \tablehead{1}{c}{b}{Data}
  & \tablehead{1}{c}{b}{SM bkg} \\ 
\hline
$ee$ + track      & 1.1  & 0 & 0.82$\pm$0.66  & 
$e^{\pm} e^{\pm}, e^{\pm} \mu^{\pm}, \mu^{\pm} \mu^{\pm} $ & 1.0  & 7.9$\pm$1.3  & 13 \\
$e\mu$ + track    & 0.3  & 0 & 0.31$\pm$0.13  & 
{\footnotesize {$\mu\mu+e/\mu$~low-$p_T$}}  & 1.0 & 0.4$\pm$0.1 & 1 \\
$\mu\mu$ + track  & 0.3  & 2 & 1.75$\pm$0.57  &  
$ee$ + track & 1.0 & 0.97$\pm$0.3 & 3 \\
$\mu^{\pm} \mu^{\pm}$ & 0.9  & 1 & 1.1 $\pm$0.4   & 
$e+e/\mu$+$e/\mu$ & 1.0 & 0.73$\pm$0.1 & 0\\
$e\tau_{had}$ + track   & 0.3  & 0 & 0.58$\pm$0.14  & 
{\footnotesize {$\mu\mu+e/\mu$~ high-$p_T$}} & 0.75 & 0.64$\pm$0.2 & 1 \\
$\mu\tau_{had}$ + track & 0.3  & 1 & 0.36$\pm$0.13  & 
$\mu e + e/\mu$  & 0.75 & 0.78$\pm$0.1 & 0\\
\hline
\end{tabular}
\caption{ Number of observed data events, compared to the number of events expected
from SM background processes, for different CDF and D\O~final states.
 The luminosity used is also given.}
\label{d0:trileptonsummary}
\end{table}

\setlength{\unitlength}{1mm}
\begin{figure}[bth]
 \begin{picture}(50,50)
 \put(-52.0,-2.0){\includegraphics{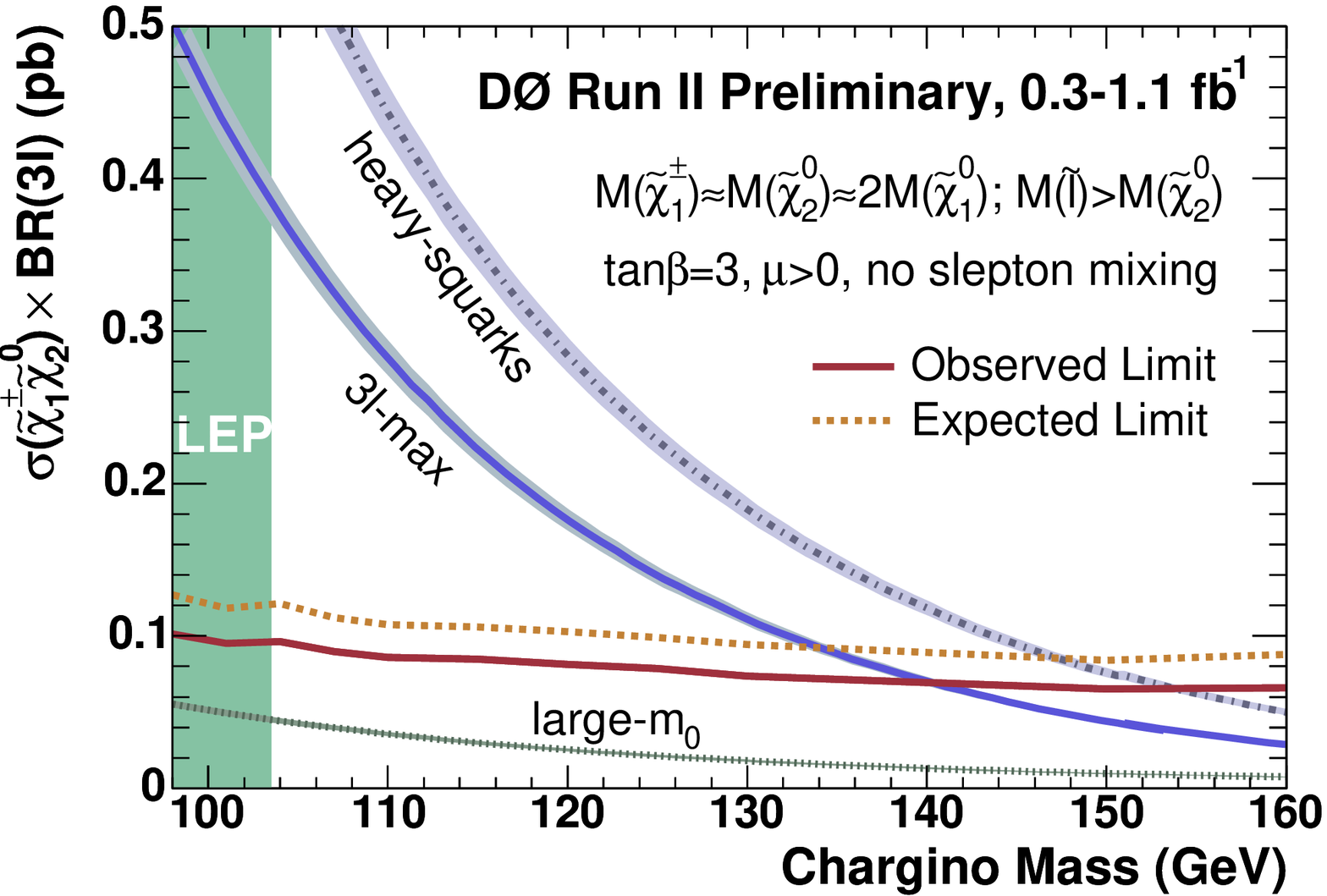}}
 \put(23.0,-2.0){\includegraphics{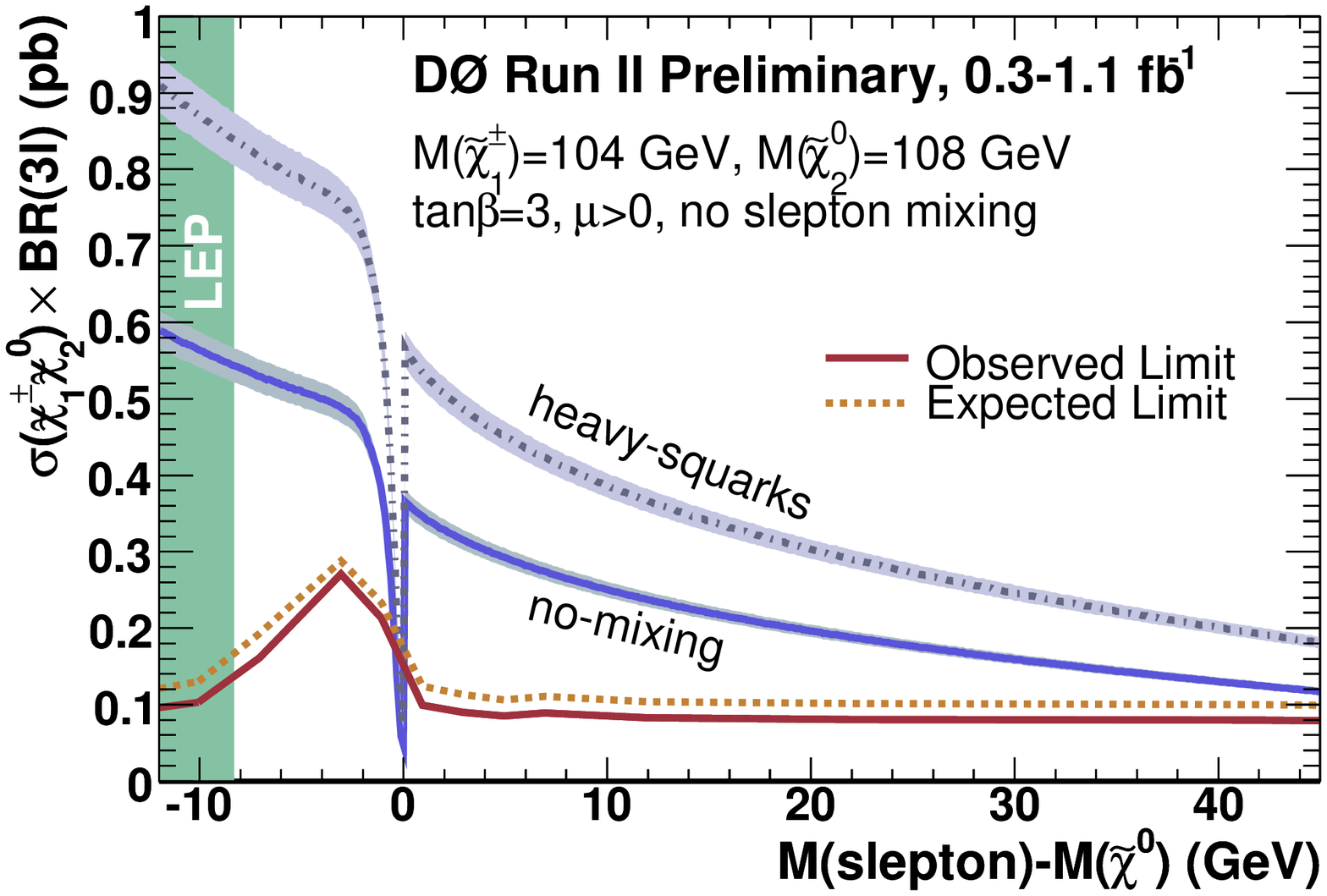}}
 \end{picture}
\caption{\footnotesize
D\O~observed limits on $\sigma ({\tilde \chi_2^0} {\tilde \chi_1^{\pm}}) \times BR (3 \lep)$
in comparison with the expectation for several SUSY scenarios 
with degenerate slepton masses, no slepton mixing, $A_0 = 0$ and $\tan \beta = 3$:
(a) as a function of chargino mass (${\tilde \chi_1^{\pm}}$) in the 3-body decay region
and (b) as a function of the slepton-neutralino ($\sle$-${\tilde \chi_2^0}$) mass difference
for $m_{{\tilde \chi_1^{\pm}}} = 104$~GeV and $m_{{\tilde \chi_2^0}} = 108$~GeV.
 PDF and renormalisation/factorisation scale uncertainties are shown as shaded bands.
 The limits are compared with the predictions of three MSSM benchmark scenarios: 
``heavy squarks'', ``$3\lep$-max'' and ``large-$m_0$'' (see text).
\label{d0rpc:chi2chi1}}
\end{figure}

 No evidence for the supersymmetry has been observed in  any of D\O~analysis.
 The SUSY expectations were optimized 
for 3-body decays via virtual $\sle^{*}$
(${\tilde \chi_2^0} \rightarrow \sle^* \lep \rightarrow {\tilde \chi_1^0} \lep \lep$), 
which are enhanced for $m_{{\tilde \chi_2^0}}, m_{{\tilde \chi_1^{\pm}}} \le m_{\sle}$
($m_0\sim90-130$~GeV, $m_{1/2}\sim170-220$~GeV).
D\O~observed limits for the combination of
$ee$+track, $e\mu$+track, $\mu \mu$~+~track and $\mu^{\pm} \mu^{\pm}$ channels
are presented in Fig.~\ref{d0rpc:chi2chi1}: (a) as a function of the ${\tilde \chi_1^{\pm}}$ mass and (b) 
of the mass difference $\Delta m = m_{\sle} - m_{{\tilde \chi_2^0}}$\footnote
{Adding $\tau$-leptons ($e+\tau_{had}+\lep$ and $\mu+\tau_{had}+\lep$ analysis)
was found to improve limits by $\sim 10~\%$ for low $\tan \beta = 3$.
For large values of $\tan \beta$, the $\sta$ becomes the lightest slepton
increasing the branching ratios to final states with 3 $\tau$'s.
The smaller $m_{\sta}$ also leads to a lower $\tau_{had}$ transverse momenta in the final state.}.
 Assuming the mSUGRA-inspired mass relation 
$m_{{\tilde \chi_1^{\pm}}} \approx m_{{\tilde \chi_2^0}} \approx 2 m_{{\tilde \chi_1^0}}$, 
the limit on $\sigma ({\tilde \chi_2^0} {\tilde \chi_1^{\pm}}) \times BR (3 \lep)$ 
can be compared with predictions from three SUSY benchmark 
scenarios~(see Fig.~\ref{d0rpc:chi2chi1}).
$W/Z$ exchange is dominant at large $m_{\sle}$ ($m_{\squark}$) masses, 
obtained by raising $m_0$ to the TeV scale and assuming scalar mass unification. 
This results in small
$BR ({\tilde \chi_1^{\pm}} {\tilde \chi_2^0} \rightarrow 3 \lep$), 
set by $W/Z$ branching ratio into leptons,
and does not allow any mass limit to be placed (``large $m_0$'' scenario).
 The trilepton final states are maximally enhanced 
for $\sle^{*}$-mediated 3-body decays 
($BR ({\tilde \chi_1^{\pm}} {\tilde \chi_2^0} \rightarrow 3 \lep) \sim 30 \%$)
for $m_{{\tilde \chi_2^0}}, m_{{\tilde \chi_1^{\pm}}} \le m_{\sle}$ (``$3\lep$-max'' scenario). 
 The ${\tilde \chi_1^{\pm}} {\tilde \chi_2^0}$ production cross section is maximal for the 
``heavy-squark scenario'',
when destructive interference between $s$ and $t$ channels is suppressed due
to the heavy squark masses. 
 This scenario is realized by relaxing scalar mass unification, when squarks are assumed
to be heavy while sleptons remain light.
 D\O~limits exclude $m_{{\tilde \chi_1^{\pm}}}  <140$~GeV for 
$\sigma\times BR \sim 0.07~$pb in ``$3\lep$-max''
and $m_{{\tilde \chi_1^{\pm}}} < 155$~GeV for 
$\sigma\times BR \sim 0.07~$pb in ``heavy-squark''
scenarios.
\setlength{\unitlength}{1mm}
\begin{figure}[bth]
 \begin{picture}(60,60)
 \put(-45.0,-3.0){\includegraphics{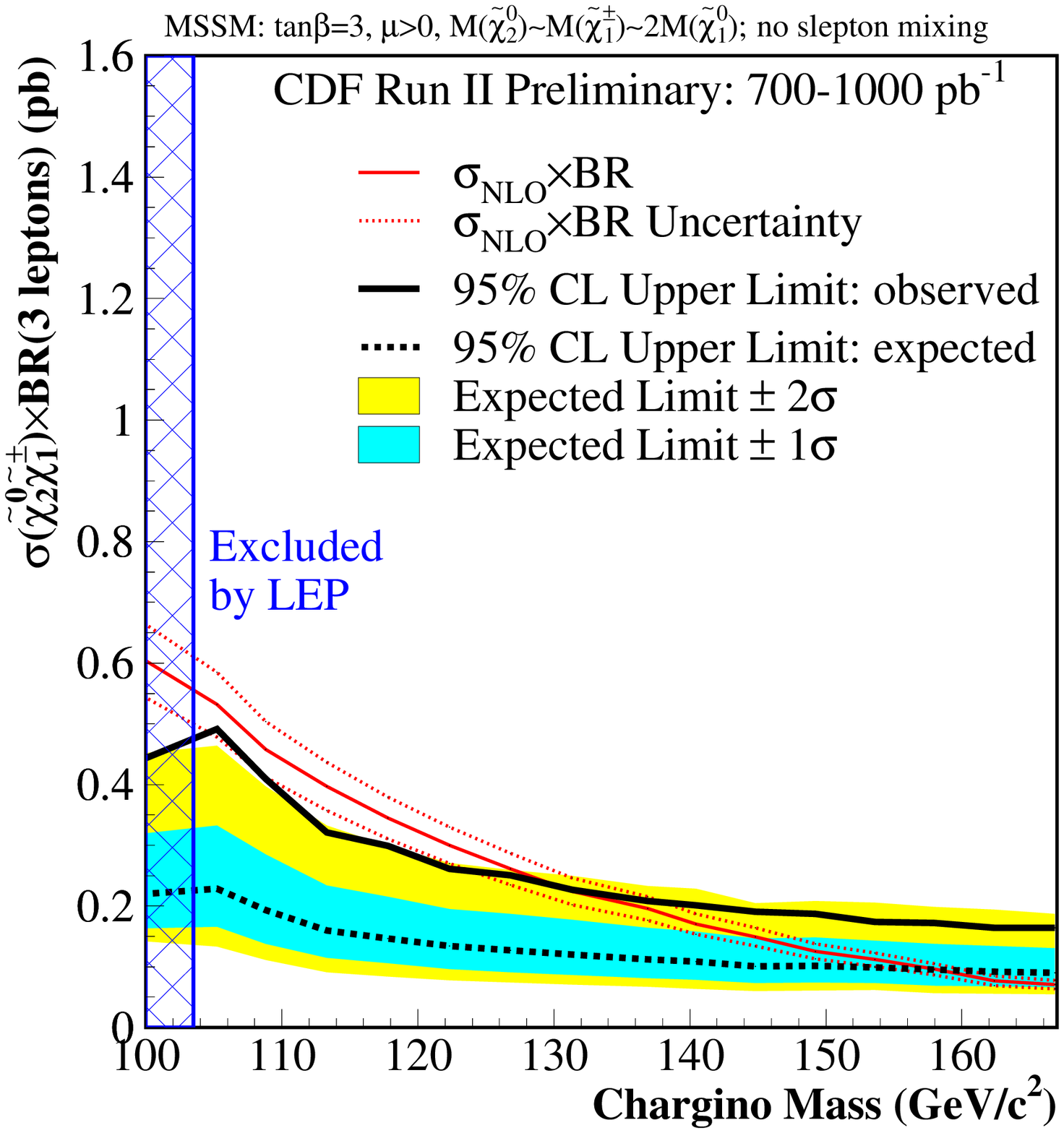}}
 \put(30.0,-5.0){\includegraphics{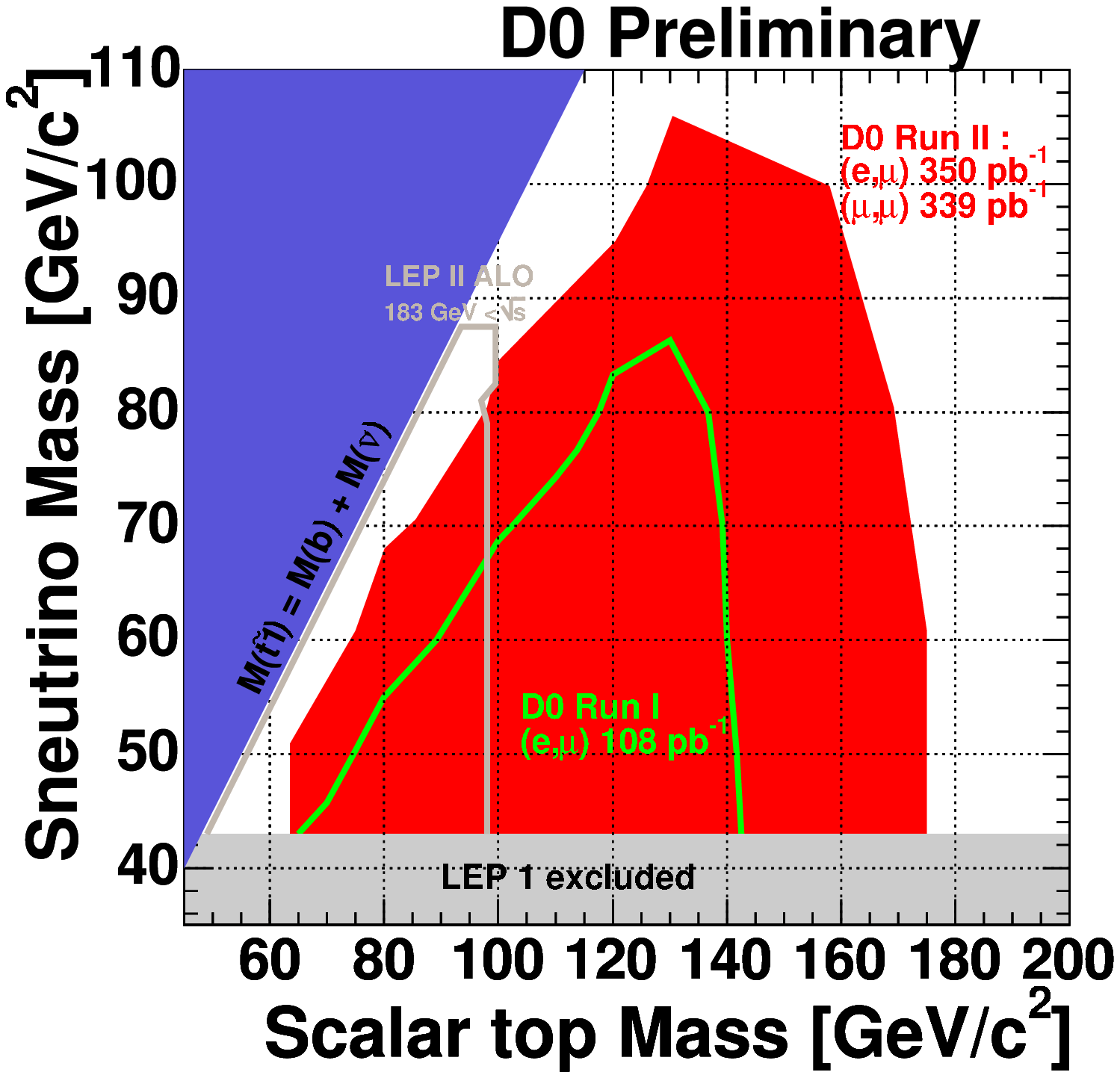}}
 \end{picture}
\caption{\footnotesize
(a) CDF expected and observed limits as a function of chargino mass (${\tilde \chi_1^{\pm}}$)
in the MSSM with degenerate slepton masses and the slepton mixing off.
(b) D\O~exclusion domains at 95~$\%$ CL in the ($m_{\stopquark },m_{\snu}$) plane
for the combination of $e \mu$ and $\mu \mu$ analysis, within the framework
of general MSSM. The results are also compared to the LEP~I, LEP~II and Tevatron Run~I
exclusion limits~\cite{d0:stop}.
\label{d0rpc:CDFchaDostop}}
\end{figure}

 Starting from ``$3~\lep$-max'' and ``heavy squark'' benchmark points
with $m_{{\tilde \chi_1^{\pm}}}(m_{{\tilde \chi_2^0}}) = 104(108)$~GeV,
$\sigma ({\tilde \chi_2^0} {\tilde \chi_1^{\pm}}) \times BR (3 \lep)$ is scanned as a function
of the slepton mass, assuming no $\sta$-mixing.
 The contribution of $\sle^{*}$-mediated 3-body decays
increases for both scenarios
with decreasing $\Delta m=m_{\sle}-m_{{\tilde \chi_2^0}}>0$, resulting in a larger 
SUSY trilepton cross section.
 For mass-degenerated $\sle$ and ${\tilde \chi_2^0}$, 
$\sigma \times BR$ vanishes since the phase space
for the $\lep$ and on-shell $\sle$ is very small and ${\tilde \chi_2^0} \rightarrow \nu \nu {\tilde \chi_1^0}$
decay dominates.
The trilepton branching ratio is further enhanced when  2-body decays via on-shell $\sle$
(${\tilde \chi_2^0} \rightarrow \sle \lep \rightarrow {\tilde \chi_1^0} \lep \lep$) dominate 
at $m_{\sle}-m_{{\tilde \chi_2^0}} < 0$.
 For $m_{\sle}-m_{{\tilde \chi_2^0}} < 6$~GeV one of the leptons from ${\tilde \chi_2^0}$
have a very low $p_T$ and only $\mu^{\pm} \mu^{\pm}$ 
analysis remains sensitive in this region.


 CDF has also completed a search in $3 \lep + \etmiss$ signature in 
different final states
depending on the leptons flavour and trigger requirements~\cite{cdftrinote}.
 Events were selected either with high-$p_T$ single lepton trigger
(indicated as ``high $p_T$'') requiring a lepton with $p_T > 18 $~GeV,
or dilepton trigger (``low $p_T$'') requiring two leptons with $p_T > 4$~GeV.
 Three analysis ($e^{\pm} e^{\pm}, e^{\pm} \mu^{\pm}, \mu^{\pm} \mu^{\pm}$) 
were based on inclusive ``high~$p_T$'' trigger and offline 
requirement that the two most energetic leptons are likesing with $p_T^{\lep } > 20(10)$~GeV.
To increase acceptance for low $p_T$ tracks the 
presence of the third lepton was not required in $\lep^{\pm} \lep^{\pm}$ events.
 The $\mu\mu + e/\mu$ final state was selected both with ``high $p_T$''
and ``low $p_T$'' triggers. In the former case 
two offline muons with $p_T^{\mu 1 (\mu 2)} > 20(5)~$GeV were selected, while
in the later case both muons were required to have $p_T^{\mu 1, \mu 2} > 5~$GeV.
 Finally, a third electron(muon) with $p_T^{\lep} > 5$~GeV was also required.
 Other trilepton channels were based on inclusive ``high $p_T$'' trigger.
 The $e+ e/\mu + e/\mu$ analysis
required one electron with $p_T^{e} > 20$~GeV,
second electron(muon) with $p_T^{\lep} > 8$~GeV and third electron(muon)
with  $p_T^{\lep} > 5$~GeV. 
 The $ee +$track channel  required two isolated electrons with
$p_T^{e1 (e2)} > 15~(5)$~GeV and an isolated track with $p_T > 4$~GeV.
 Additional cut included missing transverse energy $\etmiss > 15-20$~GeV,
depending on the final states.
 Kinematic control region were established to investigate the correct 
understanding of the SM backgrounds; whereas the signal region results were
studied only at the end. 
 After requiring channel dependent selection criteria
the observed data events agree with SM predictions within the errors, as shown in
Table~\ref{d0:trileptonsummary}.

 In the MSSM scenario with $\tan \beta=3, A_0=0$ and
$m_0 = 60$~GeV, $m_{1/2}$ ranging between 
 162 and 230~GeV and without slepton mixing
($BR ({\tilde \chi_1^{\pm}} {\tilde \chi_2^0} \rightarrow 3 \lep) \sim 60 \%$),
CDF excludes $m_{{\tilde \chi_1^{\pm}}} < 130$~GeV for 
$\sigma \times BR \sim 0.25~$pb,
as shown in Fig.~\ref{d0rpc:CDFchaDostop} (a).
 The expected limit is $m_{{\tilde \chi_1^{\pm}}} < 160$~GeV for
$\sigma \times BR \sim 0.1~$pb.
 The observed limit is approximately $2 \sigma$ away from the expected limit,
which reflects the excess of the number of observed events w.r.t 
the number of expected ones from SM background.

D\O~also searched for $p \overline{p} \rightarrow \stopquark 
\overline{\stopquark }$ production followed by a three-body
decay via a virtual chargino:
$\stopquark \rightarrow b ({\tilde \chi_1^{\pm}})^{*} \rightarrow b \lep \snu$,\footnote{
  The $BR({\tilde \chi_1^{{\pm} *}} \rightarrow b \lep \snu)$ is assumed to be 100~$\%$
and the slepton parameters has been set to obtain equal 
branching ratios into all lepton flavours.}
with either $\snu$ or ${\tilde \chi_1^0}$ being the LSP~\cite{d0note5050}.
In $R_p$-conserving scenario, the experimental signature of 
$\stopquark \overline{\stopquark }$ decays is
$2\lep+2b$-quarks+$\etmiss$, 
very similar to those expected from $t\overline{t}$ production (see Fig.~\ref{feynman:squarkgluino} (b)). 
 While the top squark production is fixed by QCD in terms of $m_{\stopquark }$,
its decay topology is solely determined by $m_{\stopquark }$ and $m_{\snu}$.
 D\O~has studied $e \mu$ and $\mu \mu$ final states.
  To maximise sensitivity close to kinematic boundary, events with
leptons transverse momenta $p_T^{\mu 1 (\mu 2)} > 8 (6)$~GeV for the $\mu^{\pm}\mu^{\mp}$
and $p_T^{e  (\mu)} > 12(8)$~GeV for the $e^{\pm}\mu^{\mp}$ channels,
$\etmiss >$15-20~GeV and satysfying
analysis dependent topological cuts were selected.
 After all selection criteria the major SM backgrounds are 
$t \overline {t}$ and $WW$ in $e\mu$ and $t \overline{t}$ in $\mu\mu$ final states.
 Since stop quarks accessible at Tevatron are lighter than $m_t$
and because the chargino, unlike W, decays via three-body mode
into a massive LSP, stop signatures are generally softer than top events.
 Therefore, shapes of the topological variable $S_T=p_T^e+p_T^{\mu}+\etmiss$ for
different $\Delta m = m_{\stopquark }-m_{\snu}$ regions and of
the scalar sum of jet transverse energies $H_T = \Sigma p_T^{jets}$
were used to discriminate
between $\stopquark \overline{\stopquark }$ signal and the SM background 
at the last analysis stage of the $e\mu$
and $\mu\mu$ analysis.
 No deviation from SM expectations were observed in $\sim 350$ pb$^{-1}$ of data. 
 Assuming lepton universality, $e\mu$ and $\mu\mu$ analysis were combined
to exclude $\stopquark$ and $\snu$ masses in the framework of general MSSM.
 The resulting plot is shown in Fig.~\ref{d0rpc:CDFchaDostop} (b)
in the ($m_{\stopquark },m_{\snu}$) plane, assuming 3-body decay 
$\stopquark \rightarrow b ({\tilde \chi_1^{\pm}})^{*} \rightarrow b \lep \snu$.
 The right edge of the exclusion contour drops at $m_{\stopquark} \sim 175$~GeV
due to the falling cross section. 
 This edge is limited by luminosity, and additional data will push the contour
to slightly higher $\stopquark $ masses.
 Because of lower $p_T$ lepton requirements, a significant extension
of the exclusion limit is achieved at small 
$m_{\stopquark }-m_{\snu}$ value.\footnote{The gap between the kinematic boundary and the left 
contour edge for low $m_{\stopquark }-m_{\snu}$ value reflects the impact of 
$\etmiss$ and lepton $p_T$ cuts.}

\section{SUSY Searches with $R$-parity violation}\label{susy:rpv}

 The search strategy for $\Rpslash$ at Tevatron
depends on the absolute value of Yukawa couplings 
$\Lambda$ ($\lambda_{ijk}, \lambda_{ijk}^{'}, \lambda_{ijk}^{''}$)
and the relative strength of the $\Rpslash$
and gauge interactions.
 The small $\Rpslash$ couplings\footnote{
For typical $\Lambda$ values $0.001 < \Lambda  < 0.01$, $\Rpslash$
interactions introduces negligible changes in production and decay
of SUSY particles, and the LSP is forced to decay within 1~cm
from primary vertex ($\Lambda >  0.001$).},
compared to electroweak interactions,
are mostly to be felt through the decay
of sparticles, otherwise pair produced via gauge couplings.
 The larger $\Rpslash$ value could manifest itself in a 
resonant production of a single slepton or squark.
 In $\Rpslash$ scenarios due to the trilinear terms, the SUSY 
particles are assumed to decay into SM particles via a single (dominant) $\Rpslash$
coupling.

 CDF and D\O~have investigated $R_p$-conserving gaugino pair production 
(${\tilde \chi_1^{\pm}} {\tilde \chi_2^0}, {\tilde \chi_1^{\pm}} \chi_1^{\mp}$),
where the produced supersymmetric particles 
decays into the LSP (assumed to be ${\tilde \chi_1^0}$), 
with $\Rpslash$ manifesting itself in ${\tilde \chi_1^0}$ decay
only\footnote{
 It is natural to assume a hierarchy of interactions, in which $\Rpslash$ 
coupling $\lambda_{ijk}$ is small compared to electroweak gauge couplings;
then $R$-parity approximately holds in decays of the heavier gauginos and 
$\Rpslash$ manifests itself in the otherwise forbidden decay of the LSP 
(assumed to be ${\tilde \chi_1^0}$ in mSUGRA).}~\cite{d0rpv,cdfrpv}.
 The decay patterns of the LSP depend on structure of $\Rpslash$
interactions. 
 In particular, if explicit $\Rpslash$ occurs through $L_i L_j \overline{E_k}$
terms the ${\tilde \chi_1^0}$ decays via
${\tilde \chi_1^0} \rightarrow \mu e \nu_e, e e \nu_{\mu} (\lambda_{121} \ne 0)$,
${\tilde \chi_1^0} \rightarrow \mu \mu \nu_e, \mu e \nu_{\mu} (\lambda_{122} \ne 0)$ and
${\tilde \chi_1^0} \rightarrow e \tau \nu_{\tau}, \tau \tau \nu_{e} (\lambda_{133} \ne 0)$
into neutrino and two charged leptons that may have different flavours 
(see Fig.~\ref{feynman:rpvsusy}).
 As the ${\tilde \chi_1^0}$ can be relatively light 
(the LEP lower mass limit $m_{{\tilde \chi_1^0}} > 50$~GeV for $\tan \beta = 5$~\cite{barbier}),
the leptons from its decay can have low $p_T$.
 Therefore, to enhance the signal sensitivity CDF and D\O~select events with 
at least 3 leptons (electrons or muons) to probe $\lambda_{121}$ and $\lambda_{122}$.
 In order for this search to be as model independent as possible,
no jet veto and only weak $\etmiss$, $\Delta \phi (\lep\lep)$ cuts and 
$\Delta \phi (\lep,\etmiss)$ are applied to remove instrumental background and cosmic
rays. This make this analysis
sensitive to any new physics in multilepton final state.
 D\O~has also studied $\lambda_{133}$ term in the final state with two electrons and $\tau_{had}$.
 The dominant SM backgrounds are similar to the $R_p$ analysis.
 In summary, no evidence for $\Rpslash$-SUSY has been found. 

\begin{table}[htb]
\begin{tabular}{ccc|ccc}
\hline
    \tablehead{1}{c}{b}{D\O~{\footnotesize($m_0$ = 1 TeV)}}
  & \tablehead{1}{c}{b}{$m_{{\tilde \chi_1^0}}$ (GeV)}
  & \tablehead{1}{c}{b}{$m_{{\tilde \chi_1^{\pm}}}$ (GeV)}
\vline
  & \tablehead{1}{c}{b}{CDF~{\bf \footnotesize($m_0$ = 250 GeV)}}
  & \tablehead{1}{c}{b}{$m_{{\tilde \chi_1^0}}$ (GeV)}
  & \tablehead{1}{c}{b}{$m_{{\tilde \chi_1^{\pm}}}$ (GeV)}\\
\hline
$ \lambda_{121}$    & 121  & 234 & 
$ \lambda_{121}$    & 101  & 185 \\
$ \lambda_{122}$    & 119  & 230 &
$ \lambda_{122}$    & 110  & 203 \\
$ \lambda_{133}$    &  87  & 167 &
      &   &    \\
\hline
\end{tabular}
\caption{ D\O~and CDF lower limits at the 95~$\%$ CL on the
 ${\tilde \chi_1^0}$ and ${\tilde \chi_1^{\pm}}$ masses obtained in the mSUGRA inspired
model. To increase the sensitivity 
D\O~combines $ee \lep$, $\mu \mu \lep$ and $ee \tau_{had}$ 
channels for each $\lambda_{ijk}$.}
\label{rpv:summary}
\end{table}
\setlength{\unitlength}{1mm}
\begin{figure}[bth]
 \begin{picture}(55,55)
 \put(-40.0,-3.0){\includegraphics{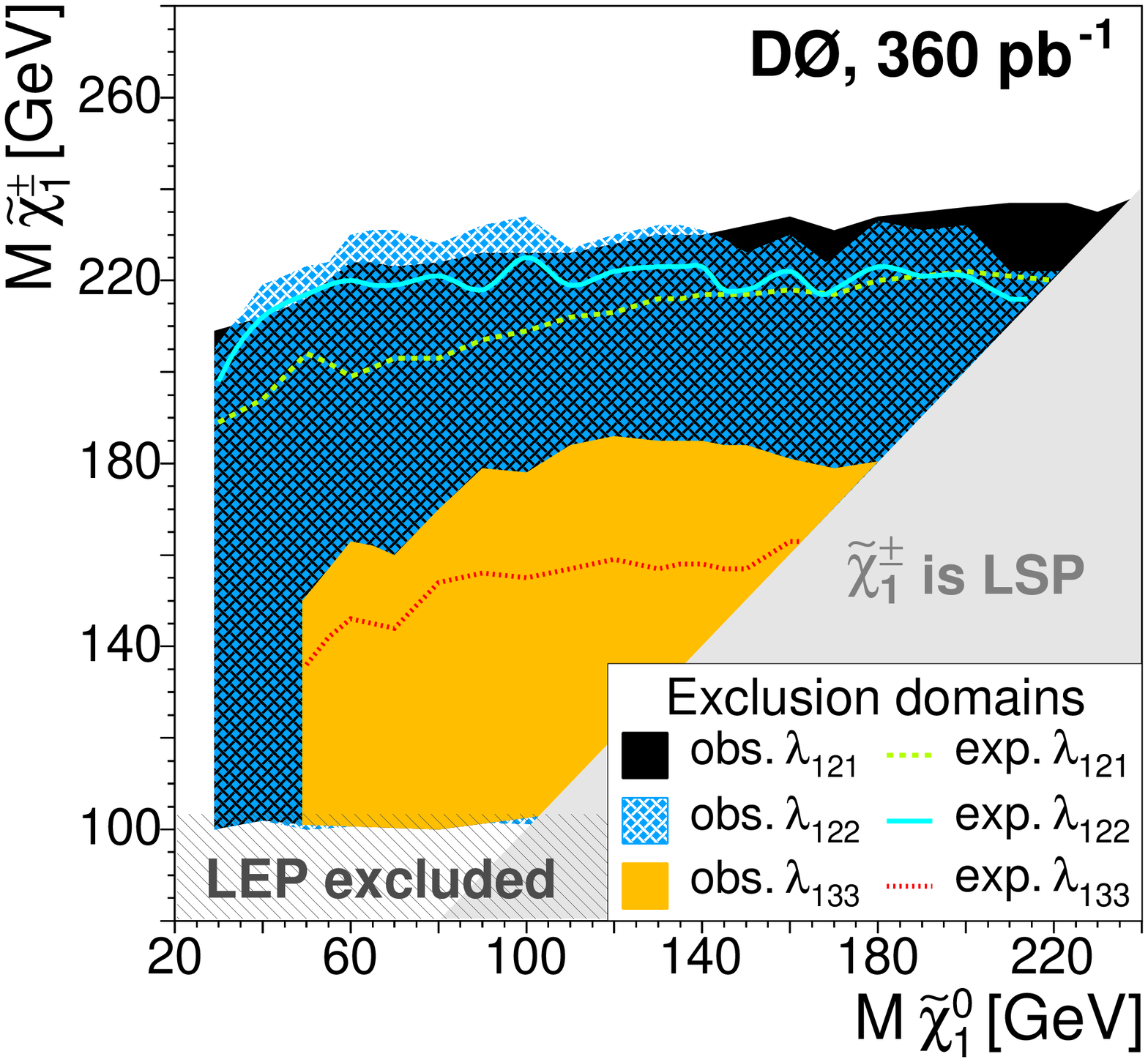}}
 \put(15.0,-155.0){\includegraphics{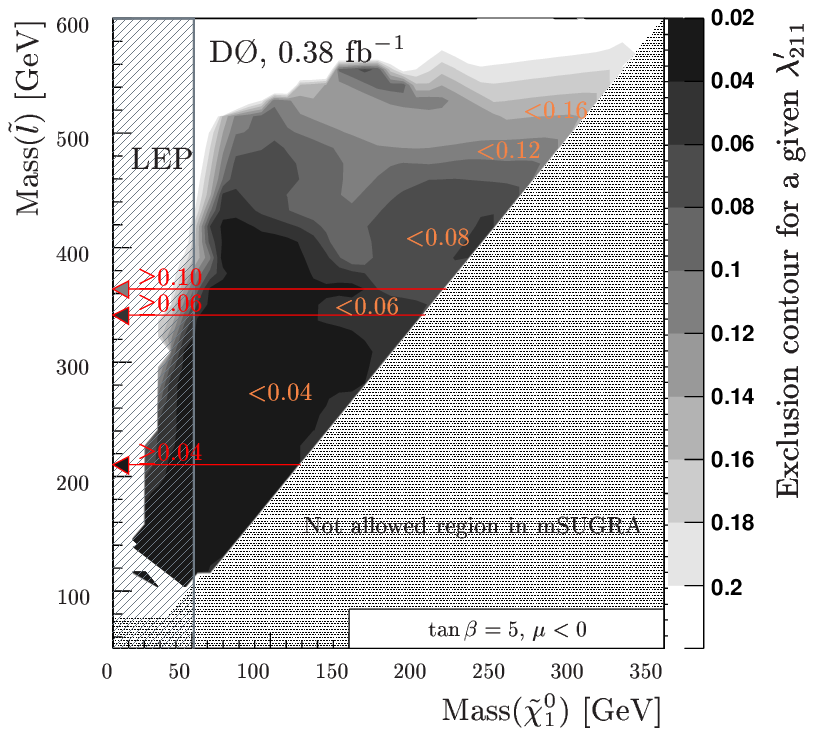}}
\put(-45.0,50.0){\bf a)}
\put(25.0,50.0){\bf b)}
 \end{picture}
\caption{\footnotesize
(a) D\O~exclusion domains at the 95~$\%$ CL in the (${\tilde \chi_1^0}, {\tilde \chi_1^{\pm}}$) mass plane
within the MSSM model 
for the $\lambda_{121}, \lambda_{122}, \lambda_{133}$ couplings (indices refer to lepton families).
(b) D\O~exclusion contour at the 95~$\%$ CL on $\lambda^{'}_{211}$ couplings within the
mSUGRA framework for $\tan \beta = 5, \mu < 0 , A_0 = 0$. The arrows indicate
limits on the slepton mass $\sle$, for a given coupling $\lambda^{'}_{211}$.
\label{d0rpV:decprod}}
\end{figure}

 Table~\ref{rpv:summary} summarises D\O~(CDF) lower mass limits obtained in the mSUGRA framework
for $m_0$ = 1~TeV\footnote{
 $m_0$ = 1~TeV corresponds to heavy sleptons, therefore exclusion limits
are valid for any slepton mass.}
 ($m_0$ = 250~GeV), $\mu > 0$, $\tan \beta = 5$, $A_0 = 0$ 
for each of the three $\lambda_{121}, \lambda_{122}, \lambda_{133}$ couplings.
D\O~also set limits for $\lambda_{133}$ at small $m_0 = 100$~GeV and $\tan \beta \sim 20$,
where the stau is the next-to-lightest supersymmetric particle,
excluding chargino (neutralino) mass below 217 (115)~GeV.
Finally, Fig.~\ref{d0rpV:decprod} (left) shows D\O~exclusion domain
in the (${\tilde \chi_1^0}, {\tilde \chi_1^{\pm}}$) mass plane 
in the MSSM scenario with heavy sfermions but
assuming no GUT-relation between $M_1$ and $M_2$.
 The cut-off at low ${\tilde \chi_1^0}$ masses $m_{{\tilde \chi_1^0}} > 30~(50)$~GeV for 
$\lambda_{121}, \lambda_{122}$~($\lambda_{133}$) 
is due to the loss of efficiency, coming from the combined effect of the coupling strength and the
LSP decay length, required  to be less than 1~cm.
All limits significantly improve previous results obtained at LEP
and at Tevatron Run~I and the most restrictive to date.

 D\O~has searched for a resonant production of single sleptons
($\smu$ or $\snu_{\mu}$) via non-zero $\lambda_{211}^{'} L Q \overline{D}$ coupling:
$q \overline{q} \rightarrow \smu~(\snu) 
\rightarrow \tilde \chi_{1,2,3,4}^0 \mu~(\tilde \chi_{1,2}^{\pm} \mu) \rightarrow 
\mu \mu \overline{q} q$ (see Fig.~\ref{feynman:rpvsusy} (c)),
assuming that it is large enough that the $\Rpslash$ decay of the LSP 
does not produce a displaced vertex~\cite{d0:slepton}.
At hadron colliders, due to the continuos energy 
distribution of the colliding partons, the resonance can be probed 
in a wide mass range of sparticle masses. 
 The $\lambda_{211}^{'}$ coupling corresponds to the first generation
of colliding partons and is not severely constrained by low
energy experiments $\lambda_{211}^{'} < 0.059 \cdot \msquark /100$~GeV,
since $\msquark$ can exceed 1~TeV for the first family~\cite{l211:limit}.
 The event topology searched is 2 isolated muons with $p_T^{\mu} > 15~(8)$~GeV,
respectively and 2 jets $p_T^{jet} > 15$~GeV.
 Using the next-to-leading muon and two leading jets the ${\tilde \chi_1^0}$ mass
is reconstructed, while a peak in the invariant mass of the two leading
muons and all found jets indicates the presence of slepton.
 No excess above the SM background was found in 375~pb$^{-1}$ of data.
 Since $\sigma (\sle) \propto (\lambda_{211}^{'})^2$, 
a lower mass limits on the resonant slepton production is set 
with respect to the coupling strength $\lambda_{211}^{'}$.
 Slepton masses below 210~(363)~GeV are excluded for $\lambda^{'}_{211}$ = 0.04~(0.10),
independent of other masses (see Fig.~\ref{d0rpV:decprod} (b)).

\section{Signature-based searches}

 CDF performed ``signature-based'' searches
for anomalous production of events containing 
$\lep\gamma$+$X$, which are rare SM processes~\cite{sig1,sig2}.

\setlength{\unitlength}{1mm}
\begin{figure}[bth]
 \begin{picture}(45,45)
 \put(87.0,-52.0){\includegraphics{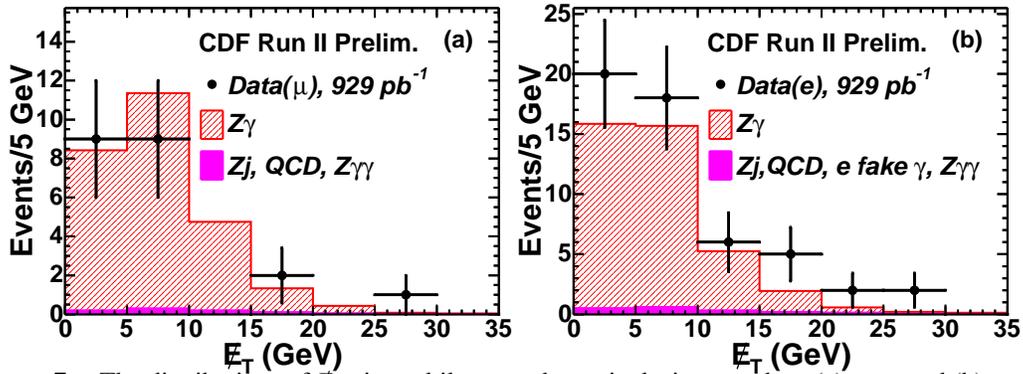}}
 \end{picture}
\caption{\footnotesize
 The distributions of $\etmiss$ in multilepton+photon inclusive searches:
(a) $\mu\mu\gamma$ and (b) $ee\gamma$. The histograms show the expected SM background. 
In the $\lep\lep\gamma$ + X sample with $\etmiss > 25$~GeV 3 events observed for the 
SM expectation of 0.6$\pm$0.1 events.
\label{cdf:signature}}
\end{figure}

 An inclusive $\lep\lep\gamma$+X sample was selected by requiring a central photon
with $E_T^{\gamma}> 25$~GeV
and a central $e,\mu$ with $E_T^{\lep}> 25$~GeV.
 The additional lepton was required to be $E_T^{\lep}> 20$~GeV.
 The dominant SM sources of $\lep\lep\gamma$ events are electroweak $Z/\gamma^{*}$
production along with $\gamma$ radiated from one of the charged particles involved in
the process.
 The 53 $ee\gamma$ and 21 $\mu\mu\gamma$ were selected in
 929~pb$^{-1}$ of data, in agreement with SM expectations
of 39.0$\pm$4.8 and 26.0$\pm$3.1, respectively.
 No $e\mu\gamma$ events were observed for an expected SM background of 1.01$\pm$0.33.
 Fig.~\ref{cdf:signature} shows the distributions of $\etmiss$
for the $\mu\mu\gamma$ and $ee\gamma$ subsamples of the $\lep\lep\gamma$ sample.
 No excess of $\lep\lep\gamma$ events with anomalous large $E_T$ or with multiple photons
were observed in CDF Run~II data.


\section{Summary and Outlook}

CDF and D\O~limits in multilepton final states presented in this paper, and based
on the luminosities between 350 and 1100~pb$^{-1}$, have significantly improved
Tevatron Run~I and LEP~II results over the large regions of SUSY parameter 
space~\cite{sig2,sig3}.
 By the end of 2006, both experiments have already collected 2~fb$^{-1}$
of data, which gives an opportunity to significantly extend searches for new physics
beyond the SM.
 In particular, recent $\chi^2$ analysis based on the present experimental results 
of the $m_W$, $sin^2 {\Theta_{eff}}$, $(g-2)_{\mu}$ and $BR (b \rightarrow s \gamma)$ 
favour relatively small chargino mass $m_{{\tilde \chi_1^{\pm}}} 
\sim 200$~GeV~\cite{senreach},
which might be within the reach of Tevatron Run~II, depending on SUSY scenario.


\begin{theacknowledgments}

 I would like to express my appreciation to the organizers for the
immensely stimulating and enjoyable conference.
 I am grateful to Jean-Francois Grivaz, Teruki Kamon and Giulia Manca
for reading and correcting this manuscript.
The conference organizers gratefully acknowledge the support of the US Department of Energy,
the US National Science Foundation, the Universities Research
Association, and the Aspen Center for Physics.

\end{theacknowledgments}

\end{document}